\documentclass[aps,pra,twocolumn,groupedaddress]{revtex4}
\usepackage{epsfig}
\usepackage{graphicx}
\usepackage{amsfonts}
\usepackage{latexsym}
\usepackage{bm}
\def\8{\infty}

\def\oh{\frac{1}{2}}

\def\d{\partial}

\def\undertext#1{\vtop{\hbox{#1}\kern 1pt \hrule}}

\def\dbyd#1#2{\frac{d#1}{d#2}}

\def\pbyp#1#2{\frac{\partial#1}{\partial#2}}

\def\be{\begin{equation}}
\def\ee{\end{equation}}
\def\bea{\begin{eqnarray} & &}
\def\eea{\end{eqnarray}}

\def\rf#1{(\ref{#1})}

\def\d{\partial}

\def\rf#1{(\ref{#1})}

\def\tlambda{\tilde \lambda}
\def\tepsilon{\tilde \epsilon}
\def\tmu{\tilde \mu}

\def\rfs#1{Eq.~\rf{#1}}

\begin{document}

%\preprint{}

\title{One dimensional gas of bosons with integrable resonant interactions}


\author{V. Gurarie}
%\thanks{}
\affiliation{Department of Physics, CB390, University of Colorado,
Boulder CO 80309, USA}


\date{\today}

\begin{abstract}
We develop an exact solution to the problem of one dimensional chiral bosons interacting via an $s$-wave Feshbach resonance. This problem is integrable, being the quantum analog of a classical two-wave model solved by the inverse scattering method thirty years ago. Its solution describes one or two branches of dressed chiral right moving molecules depending on the chemical potential (particle density). We also briefly discuss the possibility of experimental realization of such a system.
% and the extension of this model to fermions. 
\end{abstract}

%\pacs{}
%\keywords{}

\maketitle

\section{Introduction}
The advances of the last decade in the techniques of atomic physics allowed to realize a variety of exactly solvable many-body models experimentally. In particular, the Tonks-Girardeau \cite{Girardeau1960} gas, the dilute gas of repulsive one-dimensional bosons has been realized \cite{Kinoshita2004}. More generally, the Lieb-Liniger model \cite{Lieb1963} of interacting bosons in one dimensions can now be studied experimentally. 

In the field of cold atomic gases it is customary to use Feshbach resonances to control interactions between the atoms. On some level, Feshbach resonances can be thought of as simply a tool to change the interaction strength. But on a deeper level, they are a way to convert pairs of atoms into molecules whose binding energy can be controlled. When confined to one dimensions, such a system is then described by the Hamiltonian \cite{Timmermans1999}
\begin{eqnarray} \label{eq:FA}
H_F &= & \int dx~\left[\frac{1}{2m_a}  \, \partial_x \hat a^\dagger \, \d_x
\hat a+\frac{1}{2m_b} \, \d_x {\hat b}^\dagger \, \d_x \hat b + \epsilon_0 {\hat b}^\dagger
{\hat b}+ \right. \cr && \left. +
\frac{g}{\sqrt{2}} \left(\hat b~{\hat a}^{\dagger 2}+ {\hat b}^\dagger {\hat a}^2 
\right)\right],
\end{eqnarray}
Here ${\hat a}^\dagger$, $\hat a$ are the creation and annihilation operators of atoms, ${\hat b}^\dagger$, $\hat b$ are those of molecules, $m_a$, $m_b$ are their respective masses, and $\epsilon_0$ and $g$ are two parameters controlling the resonance. Throughout the paper we take both atoms and molecules to be bosons. 

The model described by \rfs{eq:FA} is unlikely to be integrable even classically \cite{Ablowitz2006}  and cannot be solved exactly. A particularly straightforward argument against quantum integrability involves calculating the amplitude of  three atom scattering  with incoming momenta $p_1$, $p_2$, $p_3$ into outgoing momenta $p_1'$, $p_2'$, $p_3'$, distinct from any permutation of $p_1$, $p_2$, $p_3$, in the first nonvanishing Born approximation. This amplitude can be verified to be nonzero, while integrability would require it to be zero \cite{Pustilnik2007}. 

 The bosonization techniques of developed in the context of Luttinger liquid theory can be applied to understand \rfs{eq:FA} \cite{Sheehy2005}. An approximate technique based on the ideas of the asymptotic Bethe ansatz can also be used \cite{Gurarie2006}. 

Yet there exist an exactly solvable model closely related to \rfs{eq:FA}. It is the model of chiral atoms and molecules interacting via a Feshbach resonance. Its Hamiltonian takes the following form
\be \label{eq:ham}
\hat H = \int dx \left[ -iu{\hat a}^\dagger \dbyd{\hat a}{x}- i v {\hat  b}^\dagger \dbyd{\hat b}{x} +
\frac{g}{\sqrt{2}} \left( \hat b~{\hat a}^{\dagger 2}+ {\hat b}^\dagger {\hat a}^2 \right) \right].
\ee
%Here ${\hat a}^\dagger$ and $\hat a$ are creation and annihilation operators of the particles of the first sort (to be referred to as ``atoms") and ${\hat b}^\dagger$ and $\hat b$ are creation and annihilation operators of the particles of the second sort (``molecules").  Both atoms and molecules are bosons. 
\rfs{eq:ham} describes the atoms which move in one direction with the velocity $u$, independent of their wave vector, and molecules which also move in one direction with the velocity $v$, also independent of their wave vector. The case of $u=v$ is degenerate, as we will see below. In what follows, it will be assumed that $u \not = v$. 

The classical version of this model is called the 2-wave model in the literature. It is known to be integrable and describes various phenomena, for example, in non-linear optics \cite{AblowitzBook2}. 

In this paper we demonstrate that this problem is also integrable quantum mechanically. We do that by employing the coordinate Bethe Ansatz and developing the exact solution of the problem defined by \rfs{eq:ham}. 

A closely related integrable model, called the 3-wave model \cite{AblowitzBook2} (which differs from \rfs{eq:ham} by having 3 chiral fields two of which can fuse into the third one) was studied and demonstrated to be integrable quantum mechanically in Ref.~\cite{Wadati1984}. This model is also interesting for the applications in quantum optics and atomic systems. Ref.~\cite{Wadati1984} did not work out the finite density behavior for the 3-wave model, which is something we do here for the 2-wave model. Studying this is thus  an interesting direction of further work.

One can argue that it is not entirely straightforward to realize the Hamiltonian \rfs{eq:ham} using real atoms and molecules. However, this Hamiltonian 
can be thought of as  an approximation to a true problem of atoms and molecules propagating in one dimension, given by \rfs{eq:FA}, if we can restrict our attention to atoms and molecules whose momentum is close to a specially chosen momentum $p_F$ (for atoms) and $2 p_F$ (for molecules). Indeed, in that case their spectrum is linear, as in
\be
\epsilon = \frac{\left( p_F + \delta p \right)^2}{2m} \approx \frac{p_F^2}{2m} + \delta p\, \frac{p_F}{m}.
\ee 
We can now interpret $p_F/m$ as the velocity $v$ the atoms in \rfs{eq:ham}. 
One can immediately see a particular difficulty with this interpretation: the mass of the molecules has to be twice that of an atom, and so are their ``Fermi" momenta. Thus the velocity of an atom and a molecule must be equal. To justify the assumption that $u \not = v$, we may have to place the atoms and the molecules on an optical lattice where the molecular and atomic matrix elements are distinct from each other. Then the effective masses of atoms and molecules no longer have to be the same. 

The interaction in the Hamiltonian \rfs{eq:ham}, controlled by the cubic Feshbach term, occurs at a point in space. In reality, however, the interactions between the atoms 
which lead to this term have a finite range which we denote $r_0$. Remembering that $r_0$ is not zero is important in what follows.

\section{The Coordinate Bethe Ansatz}

We begin by constructing a few body eigenstates of the Hamiltonian \rfs{eq:ham} and then proceed to generalize them to many body states made possible by the integrability of this problem. Then we impose the periodic boundary conditions to find the excitation spectrum of the system. In all steps we closely follow the standard techniques originally developed for the solution of the Lieb-Liniger model \cite{Lieb1963} as explained in Ref.~\cite{KorepinBook}. 

\subsection{Single atom state}
First of all, we observe that a single atom represents an exact eigenstate of the Hamiltonian \rfs{eq:ham}. This state can be written as 
\be
\int dx~ e^{i k x} {\hat a}^\dagger(x) \left| 0 \right>
\ee where $\left| 0 \right>$ is the vacuum, 
and its energy is given by $E = uk$. Indeed, all terms in the Hamiltonian besides the very first one annihilate this state. 

\subsection{Two atom states}
Now we seek the two-atom states in the following general form
\be \int dx_1 dx_2 \, \psi(x_1,x_2)\, \hat a^\dagger(x_1) \hat a^\dagger(x_2) \left| 0 \right>  + \int dy \, \phi(y)\, \hat b^\dagger(y)\left| 0 \right>
\ee
Acting on it by the Hamiltonian \rfs{eq:ham} we find the following first-quantized Schr\"odinger equation
\begin{eqnarray}\label{eq:fqs} -i u \sum_{i=1}^2 \dbyd{\psi(x_1,x_2)}{x_i} &+& \frac g {\sqrt 2} \delta(x_1-x_2) \phi(x_1) = E \psi(x_1),  \cr
-i v \dbyd{\phi(y)}{y} &+& \sqrt{2} g \, \psi(y,y) = E \phi(y). 
\end{eqnarray}
This equation has two classes of solutions. 

On the one hand, two atoms separated by some distance will forever move at equal velocity $u$ regardless of their momenta. Thus they will never interact. This is reflected in the existence of an exact eingenstate of the Hamiltonian given by $\psi(x_1,x_2) = f\left( \left| x_1-x_2 \right| \right)$, $\phi=0$, or
\be 
\int dx_1 dx_2 \, f\left(\left| x_1-x_2 \right| \right) e^{i k_1 x_1+i k_2 x_2} a^\dagger(x_1) a^\dagger(x_2) \left| 0 \right>.
\ee
Here $f(x)$ is an arbitrary function of its argument such that $f(0)=0$. The energy of this state is simply
$E=u \left(k_1+k_2 \right)$. We call this a two-atom state. 

On the other hand, two atoms located at the same point will forever be located at the same point since they move at an equal velocity. They hybridize with a molecule to form a {\sl dressed molecule} state. It is given by $\psi(x_1,x_2) = A \delta(x_1-x_2) e^{i k x_1} $, $\phi(y) = M e^{i k y}$, equivalent to  the following form
\be \label{eq:2bs}
\int dx ~e^{i k x} \left[A \, {\hat a}^{\dagger 2}(x) + M\, {\hat b}^\dagger(x) \right] \left| 0 \right>,
\ee
where $A$ and $M$ are the atom-mlecule amplitudes. Substituting these into \rfs{eq:fqs} gives
\begin{eqnarray} \label{eq:se}
uk A + \frac g {\sqrt{2}} M & = & E A, \cr
vk M + \sqrt{2} g \delta(0) A &=& E M.
\end{eqnarray}
Notice the appearance of $\delta(0)$. Treated naively, this could be interpreted as infinity. However, for the regularized model where the interactions happen at some length scale $r_0$, this term should be interpreted as $\delta(0)=1/r_0$, which is what we do in what follows. We now find from \rfs{eq:se}
\be \label{eq:energymol}
E^\pm = \frac{(v+u)k \pm \sqrt{(v-u)^2 k^2 + 4 \frac{g^2}{r_0} }}{2},
\ee
as well as
\be
\frac{M^\pm}{A^\pm} = \frac{k (v-u) \pm \sqrt{(v-u)^2 k^2 + 4 \frac{g^2}{r_0} }}{\sqrt{2}g}.
\ee
Thus the dressed molecules come in two different species, labelled by the superscripts $+$ and $-$. 
These two solution are easiest to parametrize if one introduces new variables $\lambda(k)$ which we call repidities (for reasons which will become clear later).  We define those as in
\be \label{eq:rap}
\lambda \equiv \left(v-u \right) \left[ \frac{(v-u) k \pm \sqrt{\left(v-u\right)^2 k^2 + \frac{4 g^2}{r_0}}}{4 g^2} \right].
\ee
Assuming, without the loss of generality, that $v>u$, $\lambda$ is positive if the plus sign is chosen in front of the square root and it is negative if the minus sign is chosen, thus in terms of $\lambda$, the two species of dressed molecules are straightforward to distinguish (and so, the superscript $\pm$ is not necessary for $\lambda$).  In terms of these, one can write
\be
\frac{M^\pm}{A^\pm} = \frac{2 \sqrt{2} g}{v-u} \lambda,
\ee
as well as 
\be \label{eq:freespec} E^\pm=  \epsilon_0(\lambda), \  \epsilon_0(\lambda) =  \frac{2 g^2 v \lambda}{(v-u)^2}  - \frac{u}{2 r_0 \lambda}.
\ee
Here positive $\lambda$ correspond to the choice of $+$ in the superscript  and negative $\lambda$ to the choice of $-$. 

For completeness, we also note that $k$ can be expressed in terms of $\lambda$ via
\be k = \frac{2g^2 \lambda}{ (v-u)^2} - \frac{1}{2 r_0 \lambda}.
\ee

As seen from this relation, the energy of the dressed molecule is no longer a linear function of $k$, thus the dressed molecules have a nontrivial dispersion. As a result, their velocity $\d\epsilon_0/\d k$ depends on their wave vector $k$, and unlike atoms, the molecules can catch up with each other and undergo scattering, as we will see in the next subsection. 

\subsection{Many-atomic states}

As happens in all integrable systems, more general states can be reduced to a combination of two body states. The most general eigenstate of the Hamiltonian \rfs{eq:ham} state can be written in the following form
\begin{widetext}
\begin{eqnarray} \label{eq:fullstate}
\int \prod_{n=1}^N \left[ dx_n \, e^{i k_n x_n} \left( A_n \hat a^{\dagger 2}(x_n) + M_n \hat b^\dagger(x_n)
\right) \right]
\prod_{n=1}^M \left[ dy_n \, e^{i p_n y_n} \hat a^\dagger(y_n) \right] f\left(\left| y_\alpha-y_\beta \right| \right) 
 \left| 0 \right>  \times \cr \prod_{n<m} \left[ \theta \left(x_m-x_n \right) + \theta( x_n-x_m) S_{nm}
 \right] \prod_{n,m} \left[ \theta(x_n-y_m) + \theta(y_m-x_n) S^{am}_{n} \right],
\end{eqnarray}
\end{widetext}
where $\theta(x)$ is the theta-function of its argument or $\theta(x)=1$ if $x>0$ and $\theta(x)=0$ if $x<0$. 
This state represents $N$ dressed molecules and $M$ free atoms scattering off each other. 
$A_n$ and $M_n$ are the amplitudes for the dressed molecules corresponding to the rapidities $\lambda_n$, which in turn depend on the momenta $k_n$ (and on the species of the molecule, or on whether $\lambda_n$ is positive or negative). $ f\left(\left| y_\alpha-y_\beta \right| \right)$ is an arbitrary function of all possible differences of the atomic coordinates, such that it vanishes if any two atomic coordinates coincide. $S_{nm}$ is an S-matrix representing the scattering of two dressed molecules, $n$ and $m$, off each other. Finally $S^{am}_n$ is an S-matrix for scattering between an atom and a dressed molecule, with the only index $n$ used to emphasize that it depends only on the rapidity of the molecule and not of the atom. Indeed, these $S$ matrices take the following form
\be
S_{nm} = \frac{i \left( \lambda_n-\lambda_m \right) +1 }{i \left( \lambda_n-\lambda_m \right) -1},
\, S^{am}_{n}= \frac{2 i\lambda_n -1 }{2 i \lambda_n  +1}.
\ee 
This form of the scattering matrices justifies the term ``rapidity" for the parameter $\lambda$. 
It is now a matter of a straightforward algebra to check that \rfs{eq:fullstate} is an eigenstate of the Hamiltonian \rfs{eq:ham} with the energy
\be
E = \sum_{n=1}^N \epsilon_0(\lambda_n) + u \sum_{n=1}^M p_n.
\ee

\subsection{Bethe equations}
Construction of the exact eigenstates is but the first step towards exact solution of an integrable problem using the coordinate Bethe ansatz. The next step is the imposition of the appropriate boundary conditions, the determination of the ground state energy and of the energy of the excitations above the ground states. As usual, we impose the periodic boundary conditions to arrive at the Bethe equations (here $L$ is the system size)
\be \label{eq:per}
\frac{e^{i k_j L}}{  \left[S^{am}_j \right]^M} \prod_{l \not = j} S_{jl}  =1, \ e^{i p_j L}  \prod_j S^{am}_j  =1.
\ee
$S^{am}_j$ depends on $\lambda_j$ only and is $p_j$ independent. Thus it is always possible to choose $p_j$ in such a way that
the second equation in  \rf{eq:per}
is satisfied. Then the first equation reduces to 
\be \label{eq:per1} k_j L + \sum_l \theta_{jl} - M \theta^{am}_j = 2 \pi n_j,
\ee
where $i \theta_{jl}=\ln S_{jl}$, $i \theta^{am}_j = \ln S^{am}_j$.
Following Ref.~\cite{KorepinBook} it is straightforward to prove that the solution to these equations are unique and real, and
all $n_j$ are distinct. This last claim is the consequence of the 1D ``Pauli" principle (at work here, as well as in the standard Lieb-Liniger model), which says that no two $\lambda$ can be the same, or the wave function \rfs{eq:fullstate} vanishes if $\lambda_j=\lambda_k$ for $j\not=k$ as can be checked directly. 

As a next step, we take $n_j$ to be a continuous variable $n(j)$, with $\lambda_j$ and $k_j$ becoming  functions of $n$. This gives
\be L k(n) + \sum_l \theta(\lambda(n) - \lambda(l)) - M \theta^{am} (\lambda(n))= 2 \pi n.
\ee
Here $\theta(x) = \ln \left[ (i x +1)/(i x-1)\right]/i$ and $\theta^{am}(x) =  \ln \left[ (2 i x -1)/(2 i x+1)\right]/i$

Finally, we differentiate with respect to $\lambda(n)$, introduce the function  \be
\rho = \frac 1 L \dbyd{n}{\lambda}
\ee
playing the role of the density of $\lambda$, 
and replace summation by integration to arrive at
\be \rho(\lambda)- \frac 1 {2\pi} \int d\mu \frac{2 \rho(\mu)}{(\lambda-\mu)^2 +1} =\frac 1 {2\pi} \dbyd{k}{\lambda}+ \frac{M}{2\pi L} \frac{4}{1+4 \lambda^2}.
\ee

\subsection{Yang-Yang equation}
There are two ways to make further progress in the determination of the ground state energy and the excitation spectrum of the system. One
follows explicit constructions of the excitations, by exciting a state with a particular rapidity $\lambda$, while shifting the rest of the rapidities
to accommodate the Bethe equations \rfs{eq:per1}. The second is by studying the excitations at finite temperature and then taking the limit $T \rightarrow 0$. Both methods are described in Ref.~\cite{KorepinBook}. It is technically easier to use the second approach. Although this method is well known, we go over it
briefly in the particular case of interest here. 

First we
note that in a general state  $n_j$ takes values in some subset of all possible integer numbers. We introduce $\rho_p$ as the density of $\lambda$ among the values of $n$ which are taken (``occupied"), and $\rho_h$ as the density of $\lambda$ where these values are unoccupied, with $\rho_t = \rho_p+\rho_h$ (see Ref.~\cite{KorepinBook} for the discussion on how this is done). Then we find
\be \label{eq:nonstad} \rho_t(\lambda)- \frac 1 {2\pi} \int d\mu \frac{2 \rho_p(\mu)}{(\lambda-\mu)^2 +1} =\frac 1 {2\pi} \dbyd{k}{\lambda}+ \frac{M}{2\pi L} \frac{4}{1+4 \lambda^2}.
\ee
Next we construct the energy, the entropy, and the particle number of such configuration, given by
\begin{eqnarray} E&=&L \int d\lambda \, \rho_p(\lambda) \epsilon_0(\lambda), \cr
  S& =& L \int d\lambda \left( \rho_t \ln \rho_t - \rho_p \ln \rho_p - \rho_h \ln \rho_h \right), \cr
  N &=& L \int d\lambda \, \rho(\lambda).
\end{eqnarray}
Then we minimize the thermodynamic potential $\Omega = E-T S - h N$ ($T$ is the temperature, and $h$ is chemical potential) with respect to $\rho_p$,
while remembering that the variation of $\delta \rho_p$ is related to $\delta \rho_t$ by
\be
\delta \rho_t (\lambda) = \frac 1 {2\pi} \int d\mu \frac{2}{(\lambda-\mu)^2+1} \delta \rho_p.
\ee
Following standard methods \cite{KorepinBook}, we introduce
\be \frac{\rho_h}{\rho_p} = e^{\frac {\epsilon(\lambda)}{T}}.
\ee
$\epsilon(\lambda)$ plays the role of the excitation spectrum of the system. It satisfies, as a result of the minimization of $\Omega$, 
\be \epsilon(\lambda) + \frac{T}{2\pi} \int d\mu \frac{2}{(\lambda-\mu)^2+1} \ln \left( 1+e^{-\frac{\epsilon(\mu)}{T}} \right) = \epsilon_0(\lambda)-h.
\ee
Finally, we take the limit of zero temperature, $T\rightarrow 0$. This gives the following equation
\be \label{eq:exc}
\epsilon(\lambda) - \frac{1}{2\pi} \int_{\epsilon(\mu)<0} d\mu \frac{2 \epsilon(\mu)}{\left( \lambda - \mu \right)^2+1}
 = \epsilon_0(\lambda)- h.
\ee
All of these steps are standard, with the exception of \rfs{eq:nonstad}, leading to the equation \rfs{eq:exc} which is again standard with the exception of its
nonstandard right-hand side. 
Here $h$ is the chemical potential, and the integral is taken over only the region of $\mu$ where $\epsilon(\mu) < 0$. 
Solving this equation for $\epsilon(\lambda)$ produces the excitation spectrum of the system, which is the quantity we would like to compute. Notice that the coupling constant $g$ is not explicitly present,
except through the definition of $\epsilon_0(\lambda)$ in \rfs{eq:freespec}.

If $\epsilon(\lambda)>0$, then the excitation at this $\lambda$ is a particle. If, on the other hand, $\epsilon(\lambda)<0$, then the excitation is a hole whose energy is $-\epsilon(\lambda)$. 

\section{The Excitation Spectrum}

\subsection{Dimensionless parameters}
The excitation spectrum can be found by solving the equation \rfs{eq:exc}. This can only be done numerically.  To do this in a meaningful way, let us first study the scale of the parameters involved in \rfs{eq:exc}. 

Since the interactions occur at a finite range $r_0$, we will restrict the possible values of momenta $k$ to the range \be \label{eq:range} k \in \left[-\frac{\pi}{r_0}, \frac{\pi}{r_0} \right]\ee (as if the model \rfs{eq:ham} is defined on a lattice of lattice spacing $r_0$). We would also like to make sure that the interactions are sufficiently weak so that particles moving with momenta close to $\pi/r_0$ would be close to noninteracting. This can be achieved if the $4g^2/r_0$ is much smaller than $(v-u)^2 k^2$ where $k\sim  \pi/r_0$ in \rfs{eq:energymol}. This gives
\be \frac{g^2 r_0}{(u-v)^2} \ll 1.
\ee
From now on, we adopt   this assumption. 

Second, it is convenient to rescale the rapidity $\lambda$ to simplify the expression for the energy spectrum $\epsilon_0(\lambda)$. We introduce a parameter
\be c=\frac{2 g \sqrt{r_0}}{\left| u-v \right|} \sqrt{\frac{v}{u}} \ll 1
\ee and define
\be \tilde \lambda = \lambda c.
\ee
We also introduce the dimensionless rescaled energy spectrum
\be \tilde  \epsilon = \frac{|v-u |\sqrt{r_0}}{g \sqrt{uv}} \epsilon, \   \tilde \epsilon_0 = \frac{|v-u |\sqrt{r_0}}{g \sqrt{uv}} \epsilon_0. 
\ee
The equation \rfs{eq:exc} gets simplified to
\be \label{eq:simpeq}
\tilde \epsilon(\tilde \lambda) - \frac{1}{2\pi}\int_{\tilde \epsilon(\tilde \mu) < 0} d\tilde \mu \frac{2 c \tilde \epsilon(\tilde \mu)}{(\tilde \lambda-
\tilde \mu )^2 +c^2 } = \tilde \lambda - \frac 1 {\tilde \lambda} - \tilde h.
\ee 
This equation can only be solved numerically, even in the physical limit of $c \ll 1$. 

The range of allowed momenta provides a natural cutoff for $\lambda$. \rfs{eq:rap} together with \rfs{eq:range} gives
\be \lambda \in \left[ - \frac{(v-u)^2 \pi}{2 g^2 r_0}, - \frac 1 {2 \pi} \right] \bigcup \left[ \frac 1 {2 \pi}, \frac{(v-u)^2 \pi}{2 g^2 r_0} \right] \ee
In turn, this gives for $\tilde \lambda$
\be \tilde \lambda \in \left[ - \frac{2\pi v}{u c}, - \frac{c}{2\pi} \right] \bigcup \left[ \frac{c}{2\pi}, \frac{2 \pi v}{u c} \right].
\ee
The integration range over $\tilde \mu$ in \rfs{eq:simpeq} is over these two combined intervals. 

Now we are in the position to solve \rfs{eq:simpeq} numerically.  The standard method is by interacting the relation
\be  \label{eq:iteration} \tilde \epsilon_{n+1}(\tilde \lambda) =  \frac{1}{2\pi}\int_{\tilde \epsilon_n(\tilde \mu) < 0} d\tilde \mu \frac{2 c \tilde \epsilon_n(\tilde \mu)}{(\tilde \lambda-
\tilde \mu )^2 +c^2 } + \tilde \lambda - \frac 1 {\tilde \lambda} - \tilde h.
\ee
This leads to $\tilde \epsilon_n(\tilde \lambda)$ quickly diverging to negative infinity as $n$ increases.
And indeed, the proof given in Ref. ~\cite{KorepinBook} regarding the convergence of this procedure is not applicable to \rfs{eq:simpeq}. 

Instead, we use a different technique. 
We define   a functional
\begin{widetext}
\begin{equation} Q = \frac 1 8 \int d\tilde \lambda \left( \tepsilon(\tlambda) - \left| \tepsilon(\tlambda) \right| \right)^2 - \frac 1 {16\pi} \int d\tlambda d \tmu  \frac{2 c \left(
\tepsilon(\tlambda) - \left| \tepsilon(\tlambda) \right| \right) \left(
\tepsilon(\tmu) - \left| \tepsilon(\tmu) \right| \right)  }{c^2+(\tlambda-\tmu)^2}   - \oh \int d\tlambda \left( \tlambda - \frac 1 \tlambda - \tilde h \right) \left(
\tepsilon(\tlambda) - \left| \tepsilon(\tlambda) \right| \right) 
\end{equation}
\end{widetext}

such that
\be \frac{\delta Q}{\delta \tepsilon(\tlambda)} =0
\ee is equivalent to \rfs{eq:simpeq}, up to a multiplication by $1-{\rm sign}\,\tepsilon(\tlambda)$.  Then we introduce an extra fictitious parameter $\tau$, and construct the solution to the equation
\be \label{eq:evolv} \pbyp{\tepsilon(\tlambda, \tau )}{\tau} = -  \frac{\delta Q}{\delta \tepsilon(\tlambda, \tau)} 
\ee
in the limit where $\tau \rightarrow \infty$.

This procedure allows us to compute $\tepsilon(\tlambda)$ for all such $\tlambda$ that $\tepsilon(\tlambda)<0$. 
One drawback of this procedure is that once $\tilde \epsilon(\tlambda,\tau)=0$ for some $\tlambda$ and some $\tau$, it will remain zero for larger $\tau$. 
As a result, $\tepsilon(\tlambda)$ can become ``trapped" at zero whereas it might actually be negative. We fix this problem by supplementing it with
iterations \rf{eq:iteration}. Once the initial $\tepsilon(\tlambda)$ used for iterations is close to the solution of \rfs{eq:simpeq}, subsequent iterations will not diverge. Indeed, suppose
\be \tepsilon(\tlambda) = \tepsilon_s(\tlambda) + \delta \tepsilon(\tlambda),
\ee
where $\tepsilon_s$ is the solution of \rfs{eq:simpeq}, and where $\delta \tepsilon \ll \tepsilon_s$. Then we find
\be \delta \tepsilon_{n+1}(\tlambda) =\frac 1 {2\pi} \int_{\tepsilon_s(\tilde \mu)<0} d\tilde \mu \frac{2c }{c^2+(\tilde \mu - \tlambda)^2} \delta \tepsilon_n(\tilde \mu).
\ee 
It is now fairly straightforward to prove that 
\be \int_{-\infty}^\infty d\tlambda \left( \delta \tepsilon_{n+1}(\tlambda) \right)^2 <  \int_{-\infty}^\infty d\tlambda \left( \delta \tepsilon_{n}(\tlambda) \right)^2,
\ee
which proves that the iteration procedure is not divergent.

Once we do that, we construct 
the rest of this function by using \rfs{eq:simpeq} as a definition of $\tepsilon(\tlambda)$, or
\be \label{eq:simpe1q}
\tilde \epsilon(\tilde \lambda) = \frac{1}{2\pi}\int_{\tilde \epsilon(\tilde \lambda) < 0} d\tilde \mu \frac{2 c \tilde \epsilon(\tilde \mu)}{(\tilde \lambda-
\tilde \mu )^2 +c^2 } + \tilde \lambda - \frac 1 {\tilde \lambda} - \tilde h.
\ee

\subsection{Numerical solution}
We now use this procedure to construct solutions to \rfs{eq:simpeq}. We take representative parameter values \be c =0.1, \ \frac{v}{u} = 2. \ee
We then take initial value \be \left.  \tepsilon(\tlambda, \tau)  \right|_{\tau=0} = -1.
\ee
Then we run the \rfs{eq:evolv} in steps of $d\tau=0.01$ up to $\tau=100$. The integrals are computed by discretizing the range of $\tlambda$ into $3200$ intervals. After that, we use $\tepsilon(\tlambda,100)$ as an input to the iteration
procedure \rfs{eq:iteration} where we iterate only once. This seems to be enough to generate a solution of \rfs{eq:simpeq} with a reasonable accuracy of about $10^{-3}$. The accuracy is defined as
\be \frac{ \int d\tlambda \left( \tepsilon_{n+1}(\tlambda)-\tepsilon_n(\tlambda) \right)^2}{\int d\tlambda \left( \tepsilon_{n}(\tlambda) \right)^2}.
\ee

\begin{figure}[htb]
\includegraphics[height=2in]{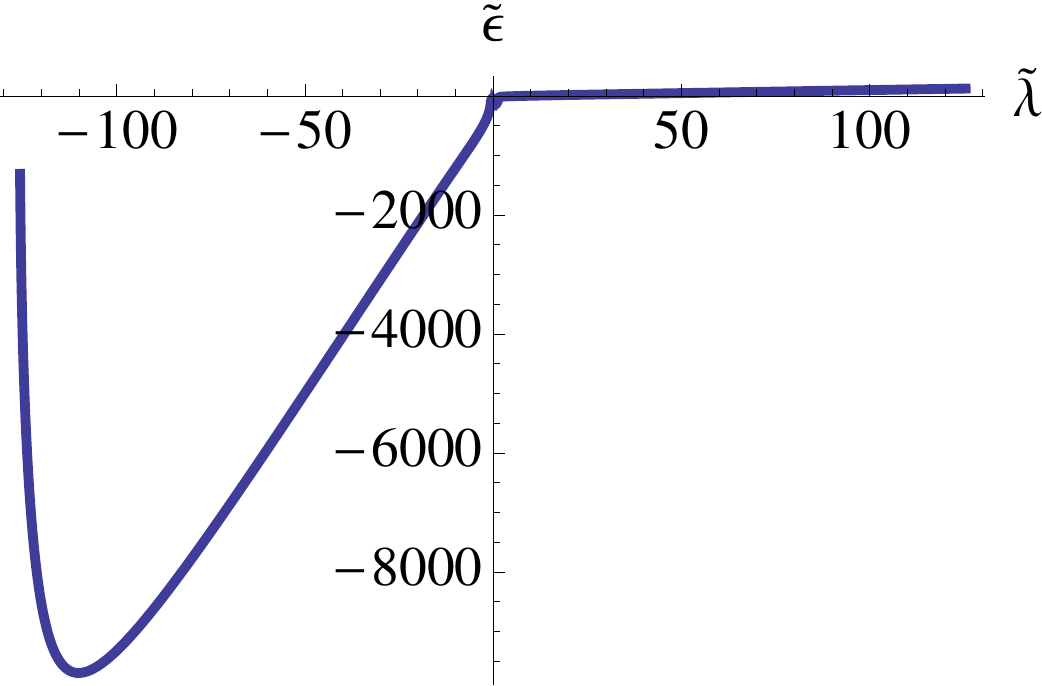}% Here is how to import EPS art 
\caption{\label{Fig1} $\tepsilon(\tlambda)$ for $c=0.1$, $v/u=2$, $h=0$.}
\end{figure}
\begin{figure}[htb]
\includegraphics[height=2in]{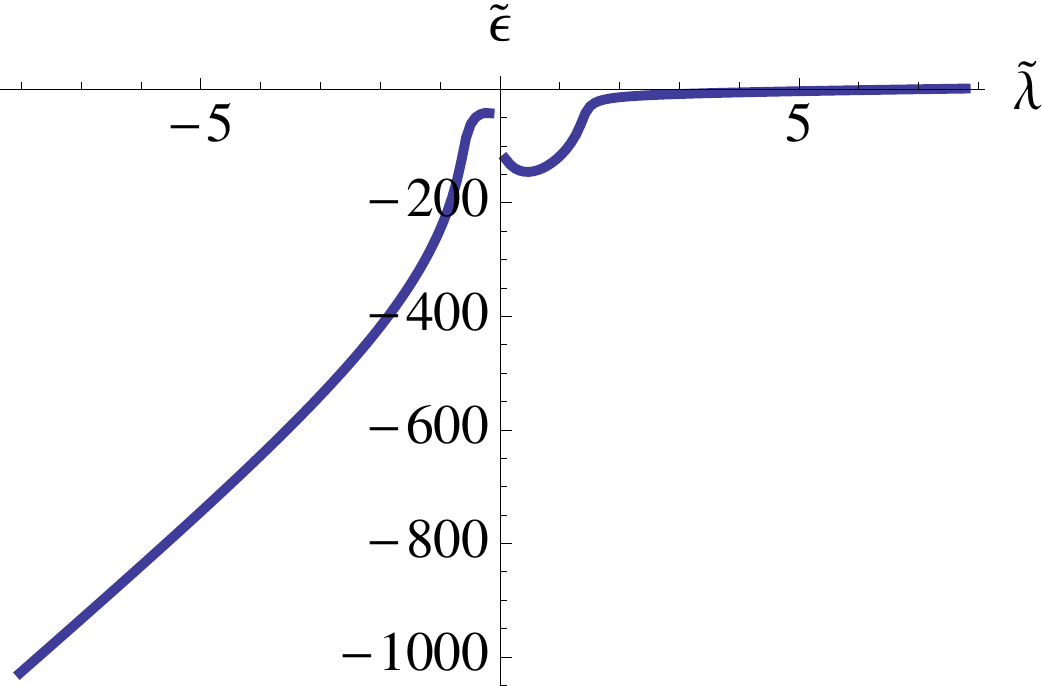}% Here is how to import EPS art 
\caption{\label{Fig2} Same figure as in Fig.~\ref{Fig1} but the region close to $\tlambda=0$ enlarged. }
\end{figure}
\begin{figure}[htb]
\includegraphics[height=2in]{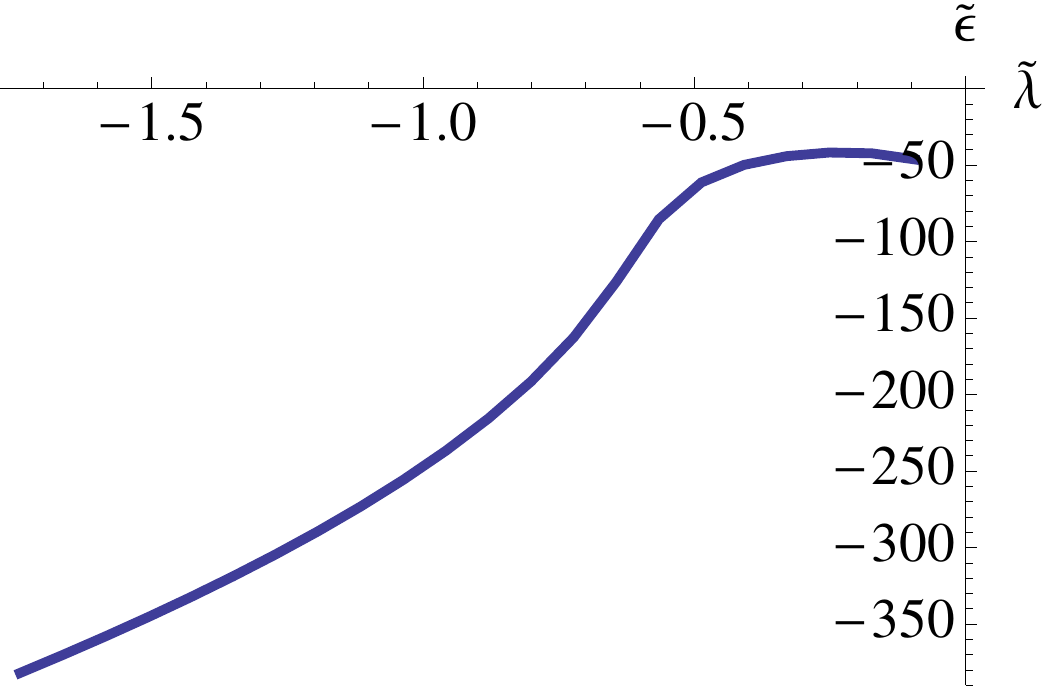}% Here is how to import EPS art 
\caption{\label{Fig3} Same figure as in Fig.~\ref{Fig1} but the region of $\tlambda<0$ close to $\tlambda=0$ enlarged. }
\end{figure}
\begin{figure}[htb]
\includegraphics[height=2in]{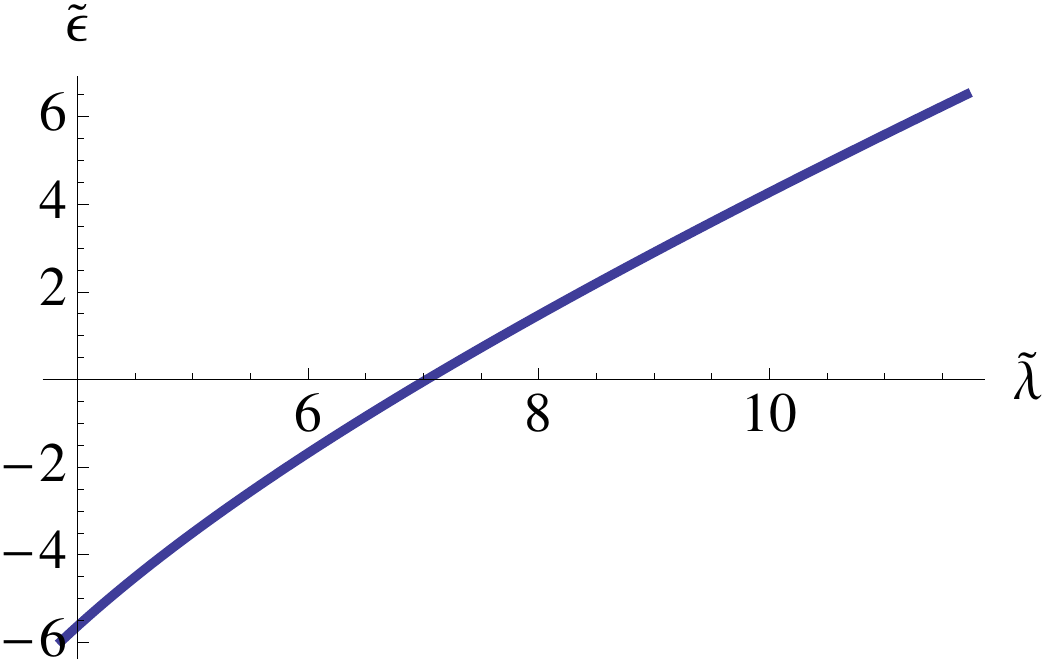}% Here is how to import EPS art 
\caption{\label{Fig4} Same figure as in Fig.~\ref{Fig1} but the region of $\tlambda>0$ close to $\tepsilon=0$ enlarged. }
\end{figure}
\begin{figure}[htb]
\includegraphics[height=2in]{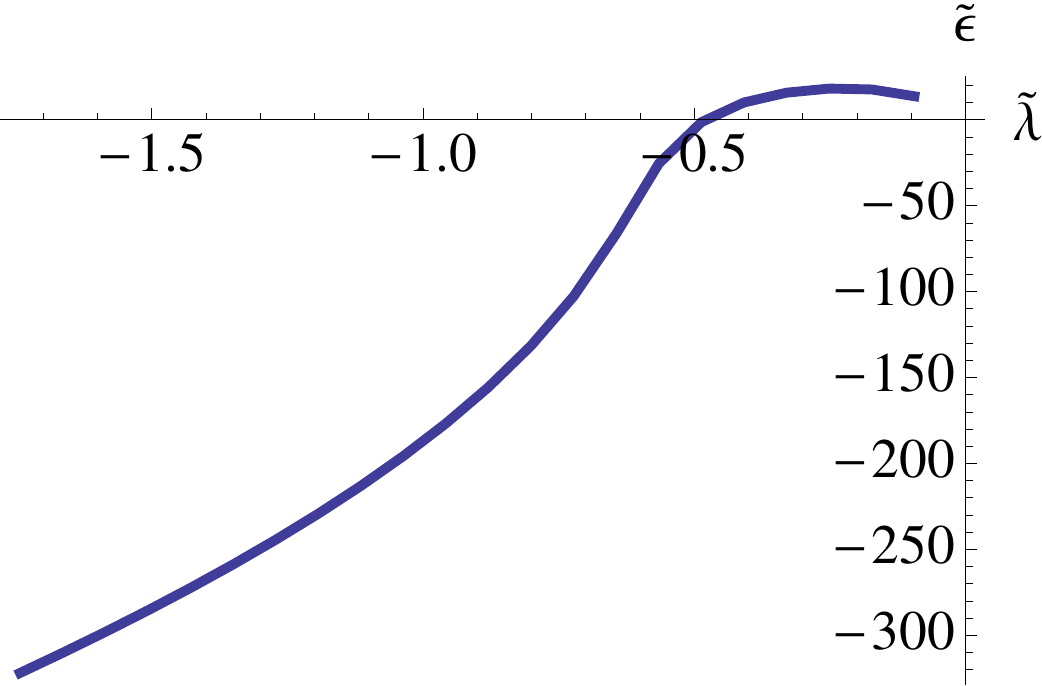}% Here is how to import EPS art 
\caption{\label{Fig5} Same figure as in Fig.~\ref{Fig3} but now the chemical potential $h=-50$. }
\end{figure}
\begin{figure}[htb]
\includegraphics[height=2in]{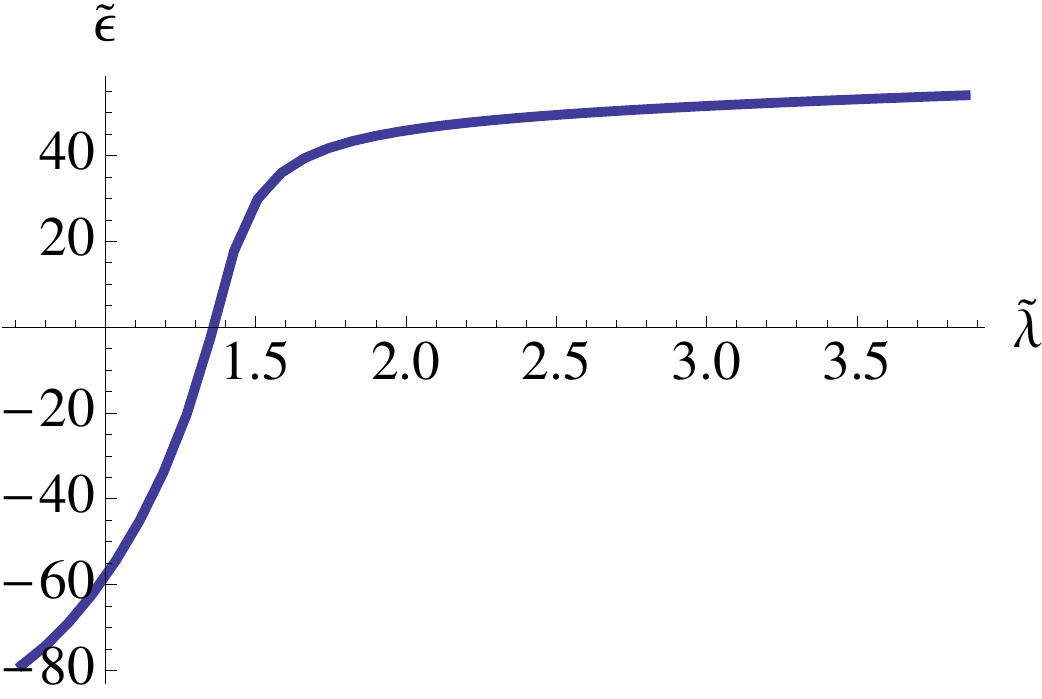}% Here is how to import EPS art 
\caption{\label{Fig6} Same figure as in Fig.~\ref{Fig4} but now the chemical potential $h=-50$. }
\end{figure}

First we illustrate the solution for $h=0$. Fig.~\ref{Fig1} shows $\tepsilon(\tlambda)$ for the entire range of $\tlambda$. Fig.~\ref{Fig2} shows the region of small $\tlambda$ enlarged. 

One sees that at $\tlambda<0$, $\tepsilon<0$. So for the branch of excitation spectrum at negative rapidities, all the excitations are holes, and they are all gapped. This is illustrated on Fig.~\ref{Fig3}. For $\tlambda>0$, $\tepsilon$ changes sign at some value of $\tlambda$. So here we have gapless right moving excitations with the linear spectrum.  This is illustrated on Fig.~\ref{Fig4}. 

By decreasing the chemical potential $h$, it is possible to make the $\tlambda<0$ molecules massless, whiling keeping the $\tlambda>0$ molecules massless as well. This is illustrated on Figs.~\ref{Fig5} and \ref{Fig6}. 

This last observation implies that the system we study undergoes a phase transition (or possibly more than one) as a function of the chemical potential (or of density), with the massless mode acquiring a gap. This phase transition looks similar to the transition observed in the problem defined by \rfs{eq:FA} in Ref.~\cite{Sheehy2005}. To elucidate the nature of this transition, it seems useful to study \rfs{eq:ham} using the bozonization techniques which we leave as a subject for future work. 

\section{Conclusions}
We have developed an exact solution to the problem of chiral atoms and molecules propagating in one dimensions with interactions controlled by a Feshbach resonance. The remaining outstanding issue is whether this Hamiltonian can be realized in a realistic cold atom experiment. 

A particular tantalizing question is whether the fermionic analog of \rfs{eq:ham} (the one where ${\hat a}^\dagger$ and ${\hat a}$ are fermionic creation and annihilation operators, and where an additional ``spin" index is necessary to make the qubic term non-zero) is integrable. If so, the development of an exact solution of such a problem would be an interesting direction of further research. The fermionic version of the 3-wave model was shown to be integrable in Ref.~\cite{Ohkuma1984}.

\acknowledgements
%\section{Acknowledgements}
The author is grateful to P. B. Wiegmann for the initial suggestion to study this problem, and to M. J. Ablowitz for discussing the classical analog of the problem defined by \rfs{eq:ham} studied here, as well as to R. Wilson for the discussions at the early stages of this project, to G. Astrakharchik and M. Hermele for useful advice concerning the numerical procedure, to L. Radzihovsky for discussing the proper interpretation of the results, and to V. Gritsev for pointing out
Refs.~\cite{Wadati1984,Ohkuma1984}. This work was supported by the NSF grant DMR-0449521.
\bibliography{bethe}

\end{document}